%% file: sigir16_cocos.tex
\documentclass{sig-alternate-05-2015}
\usepackage[
bookmarks=false
,bookmarksnumbered=true
,hypertexnames=false
,breaklinks=true
]{hyperref}
\hypersetup{
  pdfauthor={Kiseleva et al},
  pdftitle={Beyond Movie Recommendations:Solving the Continuous Cold Start Problem in E-commerceRecommendations},
  pdfsubject={2016 submission},
  pdfkeywords={H.3 [Information Storage and Retrieval]},
  pdfcreator={LaTeX with hyperref package},
  pdfproducer={pdflatex}
}
\usepackage{comment}
\usepackage{booktabs}
\usepackage{paralist}
\usepackage{xspace}
\usepackage{graphicx}
\usepackage{graphics}
\usepackage[utf8]{inputenc}
\usepackage{epsfig}
\usepackage{multirow}
\usepackage{amsmath}
\usepackage{mathtools}
\usepackage{algorithm}
\usepackage{algorithmic}
\usepackage{color}

\newcounter{todocnt}

\newcounter{latercnt}

\usepackage[square,comma,numbers,sort&compress,sectionbib]{natbib} 
\newcommand{\mypar}[1]{\medskip\noindent\textbf{#1}~}
\def\:{\hskip0pt} 
\usepackage{booktabs}
\usepackage{tabularx}
\usepackage{times}

\newcommand{\booking}{\href{http://www.booking.com}{Booking.com}\xspace}
\newcommand{\cocos}{\textsl{CoCoS}\xspace}
\newcommand{\ucocos}{\textsl{UCoCoS}\xspace}
\newcommand{\icocos}{\textsl{ICoCoS}\xspace}
\newcommand{\df}{\href{http://www.booking.com/destinationfinder.html}{Destination Finder}\xspace}

\def\sharedaffiliation{
\end{tabular}
\begin{tabular}{c}}

\begin{document}
\title{Helping To Find Best Place For Vacation: \\ Mining and Understanding  Multidimensional Situational User Profiles}
\title{Mining and Understanding Multidimensional Situational User Profiles}
\title{Utilizing Multidimensional Situational User Profiles using Multi-Criteria Rankings}
\title{Engaging Users Using Multidimensional Situational Profiles and Multi-Criteria Rankings}
\title{Utilizing Multidimensional Contextual User Profiles \\ using Multi-Criteria Rankings}
\title{Discovering and Using Contextual User Profiles\\ to Improve Destination Recommendation}
\title{Contextual User Profiles for Travel Recommendations}
\title{Discovering and Using Multidimensional Contextual User Profiles\\ to Recommend Travel Destinations}
\title{Discovering and Using Contextual User Profiles\\ for Travel Recommendations}
\title{We are not Netflix: Solving The Continuous Cold Start Problem In E-commerce Recommendations}
\title{Towards the Continuous Cold Start Problem in E-commerce Recommendations}
\title{Addressing the Continuous Cold Start Problem in E-commerce Recommendations}
\title{Contextual User Profiles for \\ Continuous Cold Start Recommendations}
\title{Optimizing Travel Destinations Based on \\ Dynamically Changing User Preferences}
\title{Contextual Ranking of Travel Destinations Based on \\ Dynamically Changing User Preferences}
\title{Ranking Travel Destinations with Contextual User Profiles}

\title{Beyond Movie Recommendations:\\ Solving the Continuous Cold Start Problem in E-commerce Recommendations}

\numberofauthors{1} 

\author{
\alignauthor 
\centerline{\begin{tabular}{@{}c@{~~}c@{~~}c@{}} 
\phantom{Mats Stafseng Einarsen\(^3\)}
& \phantom{Alexander Tuzhilin\(^1\)}
& \phantom{Mats Stafseng Einarsen\(^3\)}
\\[-2em]
Julia Kiseleva\(^1\)
& Alexander Tuzhilin\(^4\)
& Jaap Kamps\(^3\)
\\[0.5ex]
Melanie J.I. Mueller \(^2\) 
& Lucas Bernardi\(^2\)
& Chad Davis\(^2\)
\\[0.5ex]
  Ivan Kovacek\(^2\)
& Mats Stafseng Einarsen\(^2\)
& Djoerd Hiemstra\(^5\)
\\[-1ex]
\end{tabular}}
\sharedaffiliation
\affaddr{\mbox{}\(^1\)Eindhoven University of Technology, Eindhoven, The Netherlands}\\
\affaddr{\mbox{}\(^2\)Booking.com, Amsterdam, The Netherlands}\\
\affaddr{\mbox{}\(^3\)University of Amsterdam, Amsterdam, The Netherlands}\\
\affaddr{\mbox{}\(^4\)Stern School of Business, New York University, New York, USA }\\
\affaddr{\mbox{}\(^5\)University of Twente, Enschede, The Netherlands}\\[1ex]
\affaddr{\normalsize \texttt{\(^1\)j.kiseleva@tue.nl \(\quad\)  
\(^3\)kamps@uva.nl \(\quad\) 
\(^4\)atuzhili@stern.nyu.edu  \(\quad\) 
\(^5\)d.hiemstra@utwente.nl}}\\
\affaddr{\normalsize \texttt{\(^2\)\{melanie.mueller, lucas.bernardi, chad.davis, ivan.kovacek, mats.einarsen\}@booking.com}}
}


\maketitle

\newcommand{\fourrqmain}{By analyzing multi-criteria ranking data, can we automatically detect contextual user profiles that lead to better travel recommendations in terms of user engagement measures?}

\renewcommand{\fourrqmain}{Can we automatically detect contextual user profiles from multi-criteria ranking data, and will contextual profile based travel recommendations lead to an increase in user engagement within settings of \cocos?}
\renewcommand{\fourrqmain}{Can we automatically detect contextual user profiles from multi-criteria ranking data, and will contextual profile based travel recommendations lead to an increase in user engagement with continuous cold start recommendations?}

\newcommand{\fourrqone}{\textbf{RQ1:} How to characterize the continuous cold start problem in recommendation?}
\newcommand{\fourrqtwo}{\textbf{RQ2:} How to define and discover contextual user profiles taking into account the multidimensional rankings in an unsupervised setup within settings of \cocos?}
\renewcommand{\fourrqtwo}{\textbf{RQ2:} How to define and discover contextual user profiles from multi-criteria ranking data in an unsupervised setup?}
\newcommand{\fourrqthree}{\textbf{RQ3:} How to apply 
contextual user profiles for customized recommendations of travel destinations in a continuous cold start setting?}
\newcommand{\fourrqfour}{\textbf{RQ4:} How effective is our approach for real-world users of the destination finder recommendation engine?} 

\newcommand{\chap}{4.}
\renewcommand{\chap}{}

\renewcommand{\fourrqmain}{Can we automatically detect contextual user profiles and does customized ranking with these profiles improve travel search and recommendation?}
\renewcommand{\fourrqone}{\textbf{RQ\chap1:} How to characterize the continuous cold start problem in travel recommendation?}
\renewcommand{\fourrqtwo}{\textbf{RQ\chap2:} How to define and discover contextual user profiles from multi-criteria ranking data in an unsupervised setup?}
\renewcommand{\fourrqthree}{\textbf{RQ\chap3:} How to apply contextual user profiles for the ranking of travel destinations in a continuous cold start setting?}
\renewcommand{\fourrqfour}{\textbf{RQ\chap4:} How effective are contextual profiles for real-world users of the destination finder system in terms of user engagement measures?} 

\begin{abstract}
     Many e-commerce websites use recommender systems or personalized rankers to 
     personalize search results based on their previous interactions. However, a large fraction of users has no prior interactions, making it impossible to use collaborative filtering or rely on user history for personalization.  Even the most active users may visit only a few times a year and may have volatile needs 
     or different personas, making their personal history a sparse and noisy signal at best.
     This paper investigates 
     how, when we cannot rely on the user history, the large scale availability of other user interactions still allows us to build meaningful profiles from the contextual data and whether such contextual profiles are useful to customize the ranking, exemplified by data from a major online travel agent \booking.
     
     
    %
    Our main findings are threefold:
    First, we characterize the \textsl{Continuous Cold Start Problem} (\cocos) from the viewpoint of typical e-commerce applications.
    Second, as explicit situational context is not available in typical real world applications, implicit cues from transaction logs used at scale can capture essential features of situational context.
    Third, contextual user profiles can be created offline, resulting in a set of smaller models compared to a single huge non-contextual model, making contextual ranking available with negligible CPU and memory footprint.
    Finally we conclude that, in an online A/B test on live users, our contextual ranker increased user engagement substantially over a non-contextual baseline, with click-through-rate (CTR) increased by 20\%. 
    This clearly demonstrates the value of contextual user profiles in a real world application.
\end{abstract}

\input{four-sec/1-introduction}

\input{four-sec/6-related_work}

\input{four-sec/2-problem_setup}

\input{four-sec/3-general_approach}

\input{four-sec/4-applied_approach}

\input{four-sec/5-experimentation}


\section{Conclusions}
\label{foursec:conclusion}

This paper investigated the common case in e-commerce websites relying on search and recommendation to satisfy their user's needs, yet standard personalization and recommender systems rely on rich user profiles but the majority of users are new or visit highly infrequently\:---\:we face a continuous cold start recommendation problem.
We specifically studied this problem in the context of 
one of the largest travel websites, \booking, and its \df service.


Our first research question was \textsl{\fourrqone}\ 
We introduced and characterized the \textsl{Continues Cold Start Problem} (\cocos) that happens when users (\ucocos) or/and items (\icocos)
remain ‘cold’ for a long time, and can even ‘cool down’ again after some time due to some external signals. 

Our second research question was \textsl{\fourrqtwo}\ 
We presented a general framework for discovering and using contextual user profiles.
Since we work in settings of \cocos clients visit infrequently and have volatile interests, we cannot rely on historical user interactions. Mining situational profiles to which we can map an incoming user is an effective way to deal with data sparsity and changing user interests.
In principle, any contextual features can be used, including relatively shallow implicit situational context available in any online context.
Also any ratings, reviews or other multi-criteria ranking data can be used, including travel endorsements. Similar endorsement data is being used in a venue recommendation benchmark~\citep{trec:url}.
     
Our third research question was \textsl{\fourrqthree}\ 
We used the general framework for discovering contextual user profiles for the \df. As explicit situational context is not available in typical real world application, implicit cues from transaction logs used at scale can capture the essential features of situational context.  We contextualized reviews with user agent and time data. Our main goal is to determine the impact of contextual ranking, hence we used standard methods, specifically k-means for clustering and Naive Bayes for ranking. We mapped incoming users to the nearest cluster based on Euclidean distance. 
     
Our fourth research question was \textsl{\fourrqfour}\ 
We compared our contextual travel recommendations to a non-contextual ranker. This is a hard baseline corresponding to the current live system. 
Contextual user profiles can be created offline, resulting in a set of smaller models compared to the single, huge, non-contextual model, making contextual ranking available with negligible CPU or memory footprint.  We observed an increase in user engagement, with higher click-through rates (20\%) and higher clicks per user (21\%).
   
Our general conclusion is that our contextual ranking approach shows a dramatic increase in user engagement over a non-contextual baseline, clearly demonstrating the value of contextualized profiles in a real world application that suffers from \cocos.
We focused on an e-commerce setting, applicable to millions of online companies, where the continuous cold start is the rule rather than the exception.  But also in settings such as the internet search engines where interactions are frequent and rich profiles are typically available, our approach has large potential value.  The problem of fast changing content is well-known \citep{dong:towa10}. Perhaps the fraction of new users is small, yet they may be important enough to warrant extra effort, think of new users considering a search engine switch \citep{whit:char09}.

Our future work is to further investigate the following directions. First, we plan to extend the contextual space, for example using the geographical location of the user. However, this is not straightforward since simple splitting using some ontological knowledge, e.g.\ country, can lead to very skewed distributions of traffic within the contextual features and fails to capture deeper relations in the data. More generally, we plan to look into unsupervised techniques for the context discovery, over a wider range of contextual conditions including aspects of the session at hand.
Second, it is promising to extend our method of mapping incoming users to one of the discovered CUPs to a `fuzzy' mapping in which a user can be assigned to two or more CUPs. This will allow to serve a personalized ranking based on the resulting mixture weights in the model, while still maintaining online efficiency.
Third, we will look into possibilities of more efficient and accurate CUPs discovery techniques, looking also in adaptive models that take into account long term trends such as seasonal differences.

\mypar{Acknowledgments}
This work was done while the first author was an intern at \url{Booking.com}.
We thank Lukas Vermeer and Athanasios Noulas for fruitful discussions.
This research has been partly supported by STW (CAPA project, \# 11736).

\renewcommand{\bibsection}{\section*{REFERENCES}}
\bibliographystyle{abbrvnat-tweaked}
\renewcommand{\bibfont}{\raggedright\footnotesize} 
\setlength{\bibsep}{0.35\itemsep} 
\bibliography{bibliography}
\balancecolumns 
\end{document}

%% file: four-sec/1-introduction.tex
\section{Introduction}
\label{foursec:intro}

In addition to the handful of general web search engines, there are millions of online e-\:commerce websites driving the online economy \citep{estat:url}.  
Many of these e-\:commerce websites are built around personalized search and recommendations systems. Amazon.com recommends books, Booking.com recommends accommodations and destinations, Netflix recommends movies, Reddit recommends news stories and so on. 
Recommender systems predict unknown ratings based on past or/and current information about users and items, such as past user ratings, user profiles, item descriptions. If this information is not available for new users or items, the recommender system runs into the \textsl{Standard Cold Start Problem}: it does not know what to recommend until the new, `cold' user or item gets `warmed-up', i.e.\ until enough information has been received to produce recommendations. For example, which hotels should be recommended to someone who visits \booking\ for the first time? If the recommender system is based on the history of users click` in the past, the first recommendations can only be made after the user has clicked on a couple of hotels on the website. 




Several approaches have been proposed 
to deal with the cold-start problem, such as utilizing baselines for cold users~\citep{Kluver:2014:ERB:2645710.2645742}, combining collaborative filtering with content-based recommenders in hybrid systems~\citep{schein2002methods}, eliciting ratings from new users~\citep{RashidRiedls02}, promoting diversity in recommendations~\citep{Ho_wsdm_2014}, or exploiting the social network of users~\citep{sedhain2014social}. In particular, content-based approaches have been very successful in dealing with cold-start problems in collaborative filtering~\citep{schein2002methods, SaveskiMantrach14}.
However, these approaches deal explicitly with `cold' users or items, and provide a `fix' until enough information has been gathered to apply the core recommender system. Thus, rather than providing unified recommendations for `cold' and `warm' users, they temporarily bridge the period during which the user or item is `cold' until it is `warm'. This can be very successful in situations in which this warm-up period is short, and when warmed-up users or items stay warm. 

\begin{figure}[!tb]
\centerline{%
\includegraphics[width=\linewidth]{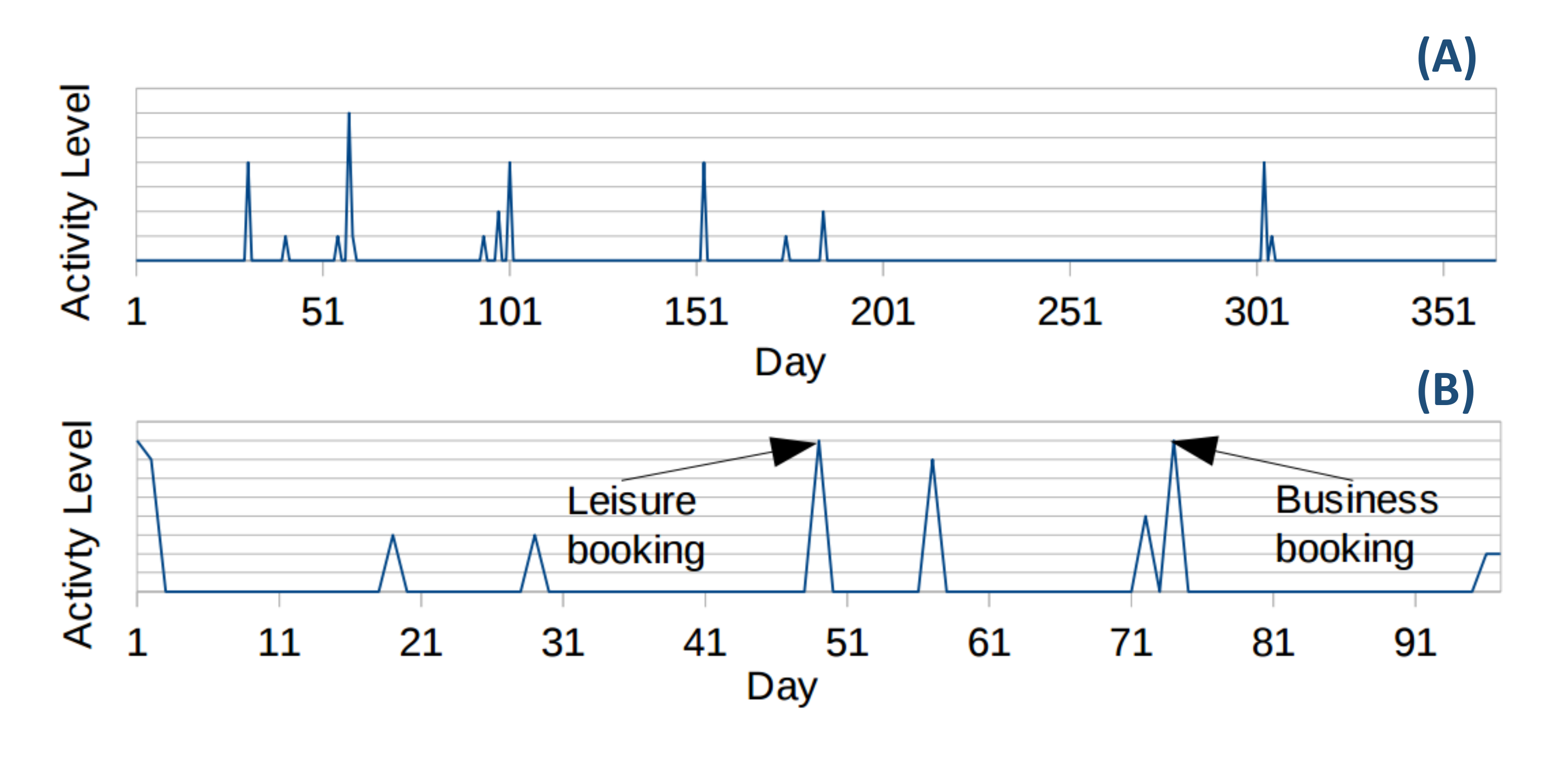}}
\vspace{-10pt}
\caption{\label{fourfig:users}Continuously `cold' users at \booking. Activity levels of two randomly chosen users over time. (A): The top user has only rare activity throughout a year. (B): the bottom user exhibits different personas by making a leisure and a business booking without much activity in between.
}
\end{figure}

\begin{figure}[!tb]
\centerline{%
\includegraphics[width=\linewidth]{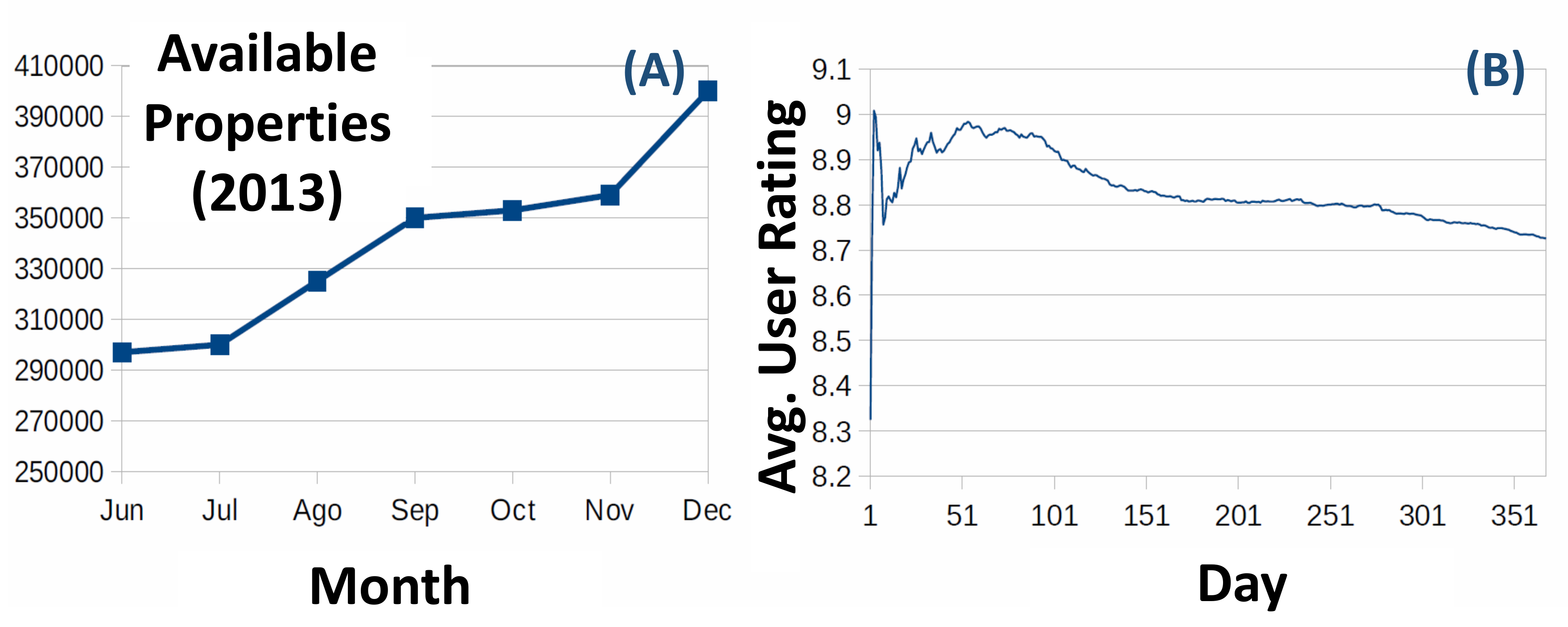}}
\caption{\label{fourfig:hotels}Continuously cold items at \booking. (A): Thousands of new accommodations are added every month. (B): The user ratings of a randomly chosen hotel change continuously over the year.}
\end{figure}


However, in many practical e-\:commerce applications, users or items remain `cold' for a long time, and can even `cool down' again, leading to the \textsl{Continuous Cold Start Problem} (\cocos). For example in \booking, many users visit and book infrequently because they have only one or two vacations per year, leading to a prolonged cold-start and extreme sparsity of collaborative filtering matrices, see Figure~\ref{fourfig:users} (A). In addition, even `long term warm' users can cool down as they change their needs over time~\citep{Kapoor_wsdm_2015}, e.g.\ coming from \booking of youth hostels for backpacking to booking of resorts for family vacations. Such `cool-downs' can happen more frequently and rapidly for users who book accommodations for different travel purposes, e.g.\ for leisure holidays and business trips as shown in Figure~\ref{fourfig:users} (B). Moreover, we have a mirror problem in the items to recommend: new items appear frequently leading to many items without prior interactions as shown in Figure~\ref{fourfig:hotels} (A) for accommodations at \booking, and items can change their characteristics as shown in Figure~\ref{fourfig:hotels} (B), making historical interactions a noisy signal.  The \cocos is ignored in the literature despite its relevance in industrial applications.  Classical approaches to the cold-start problem fail in the case of a \cocos, since they assume that users get warmed up in a reasonable time and stay warm after that.   

This paper proposes a new approach of using contextual user profiles for personalized search and recommendations 
in the context of a major online travel agent, \booking, 
in particular using the \df. 
Situational context provides powerful cues about user preferences that hold the promise to improve the quality of recommendations over the use of traditional long term interests \citep[e.g.,][]{Baltrunas_2009,Adomavicius_2005,Baltrunas_cars_2009}.
In this setup, rankings are computed based on the current context of the current visitor and the behavior of other users in similar contexts~\citep[e.g.,][]{Adomavicius_2010, Shi_cikm_2014, Hawalah_2014}. 
This type of data is readily available in most e-commerce settings. 
This approach naturally addresses sparsity by clustering users into contexts. Since context is determined on a per-action basis, user volatility and multiple personas can be addressed robustly.  

Working in a real world setting comes with specific challenges for search and recommendation systems~\citep{kohavi_kdd_2014}.  
First, in an online service, context is shallow but available at scale.  
Context can be almost anything\:---\:ranging from explicit user profiles to data about moods and attitudes\:---\:but explicit user context is typically not available in online services.  There is an abundance of situational context (day, time, device, etc) in server logs which may hold important implicit contextual cues.  Hence, although rich contextual information is not available for a large fraction of users, the large scale availability of implicit situational context may still allow us to capture essential context features.  
Second, if it's not fast it isn't working.
Due to the volume of traffic, offline processing\:---\:done once for all users\:---\:comes at marginal costs, but online processing\:---\:done separately for each user\:---\:can be excessively expensive.   Clearly, response times have to be sub-second, but even doubling the CPU or memory footprint comes at massive costs.   Hence we cannot include implicit contextual features directly or build an adaptive model for each unique user, but we can build profiles offline and map incoming users to one of the profiles at negligible online processing costs. 

We are trying to answer the following {main research question}: 
\textsl{\fourrqmain}
We breakdown of our general research problem into four specific research questions:
\begin{itemize}
\item \textsl{\fourrqone}\ 
\end{itemize}
We introduce and characterize the \textsl{Continuous Cold Start Problem} (\cocos) that happens when users or items remain `cold' for a long time, and can even `cool down' again after some time. 
\begin{itemize}
\item \textsl{\fourrqtwo}\ 
\end{itemize}
We combine multi-\:criteria ranking data with the $n$-\:dimensional contextual space in order to discover contextual user profiles.
\begin{itemize}
\item \textsl{\fourrqthree}\ 
\end{itemize}
We propose a novel approach 
exploiting contextual user profiles which are defined as \emph{‘closely connected’} regions of an $n$-\:dimensional contextual space.
\begin{itemize}
\item \textsl{\fourrqfour}\ 
\end{itemize}
We set up a large-scale online A/B testing evaluation with live traffic from \booking, and 
demonstrate how contextual travel ranking leads to a significant increase in user engagement. 

The remainder of the paper is organized as follows. 
In Section~\ref{foursec:rel_work} we discuss the most relevant prior work, and position our paper with respect to it.  
The problem setup is introduced in Section~\ref{foursec:problem_setup}.
As our approach is generally applicable to any multi-criteria ranking data associated with standard contextual information from web logs, Section~\ref{foursec:framework} outlines our approach as a general framework for discovering and using contextual user profiles.
Next, in Section~\ref{foursec:apply_framework}, we detail the specific application to our online travel agent service.  In Section~\ref{foursec:experiment}, we describe the results of the online evaluation of the approach in an A/B test with live traffic.
Finally, Section~\ref{foursec:conclusion} concludes our work in this paper and highlights future directions.

%% file: four-sec/6-related_work.tex
\section{Background and Related Work}
\label{foursec:rel_work}

In this section, we review related work in the following two areas.
First, we summarize previous work on the attempts to solve \cocos. Second, we review
approaches to build situational recommendations. 

\subsection{Cold Start Problem}
\label{foursec:rel_work_cold_start}
In classical formulations of Recommender Systems (RS), the recommendation problem relies on  \emph{ratings} ($R$) as a mechanism of capturing user ($U$) preferences for different items ($I$). The problem of estimating unknown ratings is formalized as follows: $F:U \times I \rightarrow R$. Due to practical applications, RS have been an expanding research area 
since the first papers on collaborative filtering in the 1990s~\citep{ResnickRiedl94, ShardanandMaes95}. Many different recommendation approaches have been developed since then, in particular content-based and hybrid approaches have supplemented the original collaborative approaches~\citep{AdomaviciusTuzhilin05}. For instance, RS based on latent factor models have been effectively used to understand user interests and predict future actions~\citep{Agarwal_wsdm_2010,Agarwal_kdd_2010}. Such models work by projecting users and items into a lower\:-\:dimensional space, thereby grouping similar users and items together and subsequently computing similarities between them. This approach can run into data sparsity problems and into \cocos  when new items continuously appear. 
Although, to our knowledge, the \cocos as defined in this work has not been directly addressed in the literature, several approaches are promising. 

\citet{Tang:2014:ECB:2645710.2645732} propose a context-aware recommender system, implemented as a contextual multi-armed bandits problem. 
Although the authors report extensive offline evaluation (log based and simulation based) with acceptable CTR, no comparison is made from a cold-start problem standpoint.

\citet{sun2012dynamic} explicitly attack the user volatility problem. They propose a dynamic extension of matrix factorization where the user latent space is modeled by a state space model fitted by a Kalman filter. Generative data presenting user preference transitions is used for evaluation. Improvements of RMSE when compared to time SVD \citep{koren2010collaborative} are reported. Consistent results are reported in \citep{chua2013modeling}, after offline evaluation using real data.
 
\citet{Tavakol:2014:FMD:2645710.2645739} propose a topic driven recommender system. At the user session level, the user intent is modeled as a topic distribution over all the possible item attributes. As the user interacts with the system, the user intent is predicted and recommendations are computed using the corresponding topic distribution. The topic prediction is solved by factored Markov decision processes. Evaluation on an e-\:commerce data set shows improvements when compared to collaborative filtering methods in terms of average rank.

This paper builds on our initial discussion of the cold start recommmendation problem in ecommerce practice \citep{bern_cont15}, and extends the initial experiments recommendation \citep{Kiseleva_sigir_industry_2015} by looking at ways to exploit the implicit context for increasing the effectiveness of travel recommendations in the real-world setting of \booking.

\subsection{Context-Aware Recommendations}
\label{foursec:rel_work_context}

The radical departure from classical, two-\:dimensional RS is context\:-\:aware recommendation system (CARS)~\citep{gediminasadomavicius2010}, which attract an increasing attention in academic work \citep{Shi_cikm_2014,Hariri_2012,Hariri_2013}.  Rating prediction in CARS relies primarily on the information of \emph{how} (which rating, e.g.\ a user giving `3' of `5' stars to an item) and \emph{who} (which user, e.g.\ gender, mood or nationality) rated \emph{what} (which item, e.g.\ movie, news article, or hotel). This additional information is called \emph{context}.  The general formulation of CARS rating prediction takes into account the context dimension $C$ as follows~\citep{gediminasadomavicius2010}:
\begin{equation}
\label{eq:cars}
F:U\times I \times C \rightarrow R.
\end{equation}

Defining context is an important research question in itself. The structured definition of context was introduced in~\citep{Cao_cikm_2010}. \emph{Multidimensional context} $C$ is defined as a group of contextual feature-category pairs:
\begin{equation}
 C=\{(F_n:\{v_m\}_{m=1}^M)\}_{n=1}^N, 
 \label{eq:context}
 \end{equation}
where $F_n$ are contextual features, and $v_m$ are categories for $F_n$. For example, the contextual feature \emph{location} has the contextual categories `USA', the `Netherlands' etc.  \emph{Contextual categories} are often predefined by taxonomies~\citep{Bao_2012, Zhu_cikm_2012, Hawalah_2014}. Alternatively, an unsupervised technique is used to discover contextual information~\citep{kiselevalpc_www13, Ma_www_2012}.  Moreover, context discovery can be formulated as an optimization problem~\citep{kiseleva_dddm_2013} or a feature selection problem~\citep{turny2002_discovery, Blanca_2011}.


Incorporating contextual information into CARS can be viewed as a separate area of research, and can be classified into three groups: \emph{pre-\:filtering, post-\:filtering and contextual modeling}~\citep{gediminasadomavicius2010}.
%
In the pre-\:filtering approach, contextual conditions are projected on the items, thereby essentially reducing the problem to a classical RS problem.
\citet{Adomavicius_2005} introduce a multidimensional approach taking various contextual aspects into account in collaborative filtering. They use a reduction based approach mapping a three-\:dimensional 
prediction function (of Equation~\ref{eq:context}) to a two-dimensional one.
\citet{Baltrunas_2009,Baltrunas_cars_2009} introduce item splitting for dealing with context by generating new items, where context sensitive items are duplicated and the ratings divided over the respective contextual conditions, reducing it to a classical RS problem.  This approach is expanded by \citet{Baltrunas_2014} and evaluated on synthetic and real world data sets.

Contextual information is initially ignored for post-filtering approaches, which also can be referred to as contextualization of the recommendation output~\citep{pann:expe09}. The ratings are predicted using any traditional two dimensional RS set-up on the entire data. Then, the resulting set of recommendations is adjusted (contextualized) for each user using the contextual information.

A common context modeling approach is to use contextual information to expand the feature set, thus treating context as a predictive feature. For example, \citet{steffenrendle} proposed a novel approach applying Factorization Machines~\cite{Rendle_2009} to model contextual information and provide context-aware rating predictions, \emph{using context explicitly} specified by a user to expand \emph{the set of predictive features}.

Tensor Factorization, which is a generalization of Matrix Factorization, allows a flexible and generic integration of contextual information by modeling the data as a User-Item-Context \emph{N-}dimensional tensor instead of the traditional 2D~\cite{Karatzoglou_2010,Shi_2012}.
In terms of an interactive system, the paper~\cite{palmisano2008} has shown that it was useful to consider the history of user interactions, more specifically changes in these entities. In the paper~\cite{Hariri_2012}, a co-occurrence analysis is used to mine the top frequent tags for songs from social tagging web sites, and topic modelling is used to determine a set of latent topics for each song.  Recently, more techniques for context modeling were developed~\cite{Santos_2013,Hariri_2013,Tang_2013}.



In multi-criteria RS~\cite{Adomavicius_2007, Adomavicius_2010, Lakiotaki_2011} (MCRS) the rating function has the following form: 
\begin{equation}
\label{eq:mcrs}
F:U \times I \rightarrow r_0 \times r_1\dots \times r_n.
\end{equation}
The \emph{overall rating} $r_0$ for an item shows how well the user likes this item, while criteria ratings $r_1,\dots,r_n$ provide more insight and explain which aspects of the item she likes.
MCRS predicts the overall rating for an item  based on the past ratings, using both overall and individual criteria ratings, and recommends to users the item with the best overall score. According to~\cite{Adomavicius_2007}, there are two basic approaches to compute the final rating prediction in the case when the overall rating is known.
First, in \emph{similarity based} approaches, the similarity between users is calculated based on their detailed ratings (e.g.\ Euclidean distance, Chebyshev distance, or Pearson correlation).
Second, in \emph{aggregation function based} approaches, we exploit the assumption of a relationship between the overall and the criteria ratings, $r_0=f(r_1, \dots,r_k)$ (e.g. multiple linear regression techniques can be used).
These two approaches have been significantly improved in~\citep{Jannach_2012} by using Support Vector regression and combining user- and item-based regression models with a weighted approach. \citet{Liu_2011} assumed that the overall rating highly correlates with criteria ratings that are particularly significant for individuals. 


RS methods are not easy to apply for large scale industrial applications. A large scale application of an unsupervised RS is presented in \citep{Hu:2014:SLT:2623330.2623338}, where the authors apply topic modeling techniques to discover user preferences for items in an online store. They apply Locality Sensitive Hashing techniques to overcome performance issues when computing recommendations.


\begin{figure*}[!tb]
\centerline{
\includegraphics[width=0.75\textwidth]{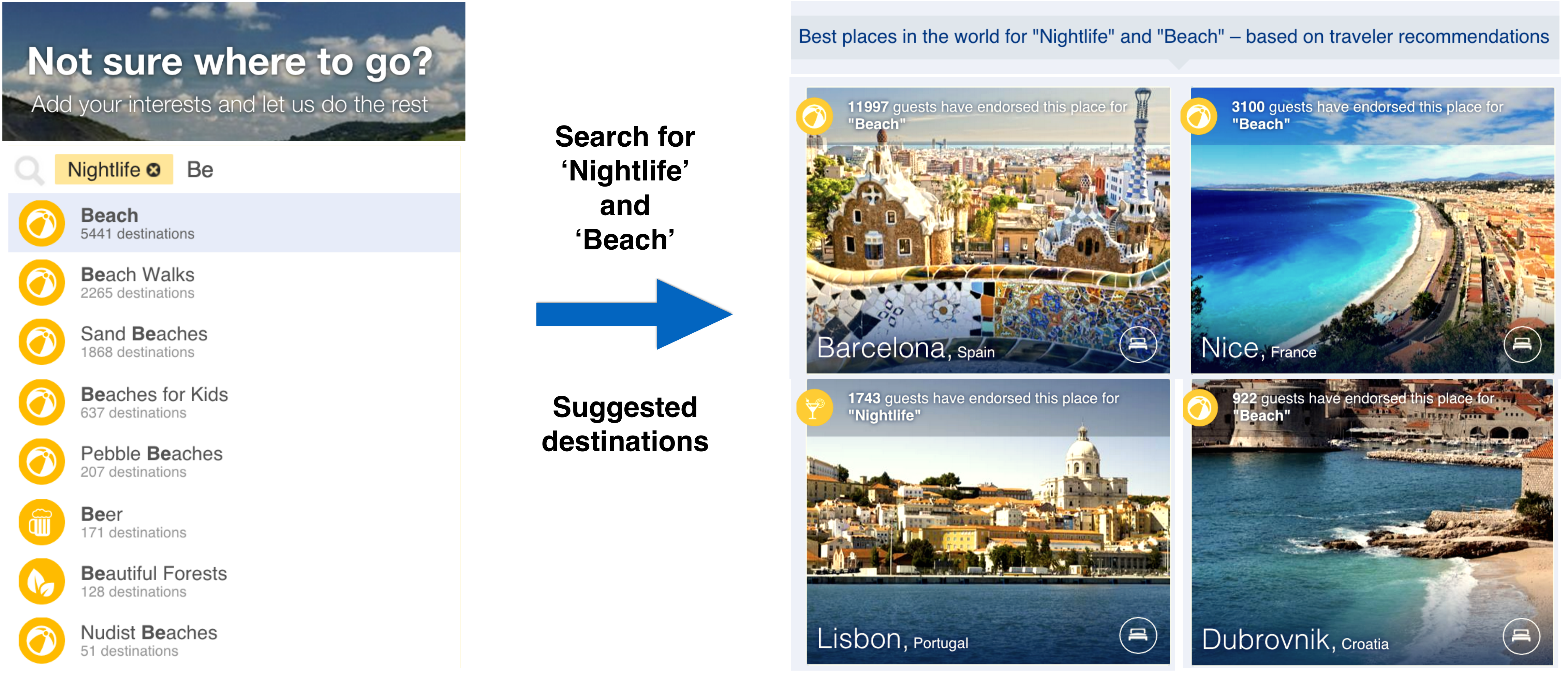}}
\caption{Example of \df use: a user searching for `Nightlife' and `Beach' obtains a ranked list of recommended destinations (top 4 are shown).}
\label{fourfig:use_cases}
\end{figure*}

\medskip
To summarize, the key distinction of our work compared to previous efforts is twofold: First, we introduce new Continuous Cold Start (\cocos) settings that is common in e-commerce.
Second, we propose the discovery of contextual user profiles (CUPs) within a \cocos setting.  CUPs are used both to build customized context-aware rankers (which can be done offline), and to map incoming users to the closest contextual user profile to provide contextual recommendations.

%% file: four-sec/2-problem_setup.tex
\section{Problem Setup}
\label{foursec:problem_setup}
In this section we will study our \textsl{\fourrqone}\
First, we characterize the \textsl{Continues Cold Start Problem} (\cocos) in Section~\ref{foursec:cocos}.  Second, we introduce a \booking service \df that `suffers' from \cocos in Section~\ref{foursec:dest_finder}.  It will be our platform for experimentation in the remainder of the paper.



\subsection{Characterizing Continuous Cold Start}
\label{foursec:cocos}
\cocos can in principle arise on both the user side and the items side. We characterize it using the following four features: \textbf{S}: data \emph{sparsity}, related to the original cold-start problem; 
\textbf{V}: \emph{volatility}, or the degree of variation in the object of interest;
\textbf{I}: object \emph{identity}, due to different technical~\citep{Montanez_cikm_2015} or law regulation related problems complicating correct identification;
\textbf{P}: `\emph{personas}', or the different types of behavior expressed by one user in different situations.

The \textsl{User Continuous Cold Start Problem} (\ucocos) can be characterized by:
\begin{itemize}
 \setlength{\itemsep}{0pt}
 \setlength{\parskip}{4pt}
 \setlength{\parsep}{4pt}
\item \textbf{S}: new or rare users;
\item \textbf{V}: users' interests change over time;
\item \textbf{I}: a failure to match data from the same user;
\item \textbf{P}: users have different interests at different, possibly close-by points in time.
\end{itemize}
New users arrive frequently as shown in Figure~\ref{fourfig:users}(A), or  may appear new when they do not log in or use a different device so we would fail to match their identity. Some websites are prone to extreme sparsity in user activity when items are purchased only rarely, such as travel, cars etc. Most users change their interests over time (volatility), e.g.\ movie preferences evolve, or travel needs change. On even shorter timescales, users have different personas. Depending on their mood or their social context, they might be interested in watching different movies. Depending on the weather or their travel purpose, they may want to book different types of trips as presented in Figure~\ref{fourfig:users} (B).

Similarly we characterize \textsl{Item Continues Cold Start Problem} (\icocos):
\begin{itemize}
 \setlength{\itemsep}{0pt}
 \setlength{\parskip}{4pt}
 \setlength{\parsep}{4pt}
\item \textbf{S}: new or rare items;
\item \textbf{V}: item properties or value change over time;
\item \textbf{I}: a failure to match data from the same item;
\item \textbf{P}: an item appeals to different types of users.
\end{itemize}
New items appear frequently in e-\:commerce catalogues, as shown in Figure~\ref{fourfig:hotels} (A) for accommodations at \booking. Some items are interesting only to niche audiences, or sold only rarely, for example books or movies on specialized topics. Items can be volatile if their properties change over time, such as a phone that becomes outdated once a newer model is released, or a hotel that undergoes a renovation. Figure~\ref{fourfig:hotels} (B) shows fluctuations of the review score of a hotel at \booking. Some items have different `personas' in that they target several user groups, such as a hotel that caters to business as well as leisure travellers. When several sellers can add items to an e-\:commerce catalogue, or when several catalogues are combined, correctly matching items can be problematic so we run into an item identity problem.

\subsection{Optimizing Destination List within CoCoS}
\label{foursec:dest_finder}

To motivate our problem set-up, we introduce a \booking service which allows to find travel destinations based on users' preferred activities: the \df.
Consider a user who knows what activities she wants to do during her holidays, and is looking for travel destinations matching these activities. This process is a complex exploratory recommendation task in which users start by entering activities in the search box as shown in Figure~\ref{fourfig:use_cases}. The service returns a ranked list of recommended destinations~\citep{Kiseleva_sigir_industry_2015xxx}. 

\begin{figure}[!tbp]
\centerline{%
\includegraphics[width=\linewidth]{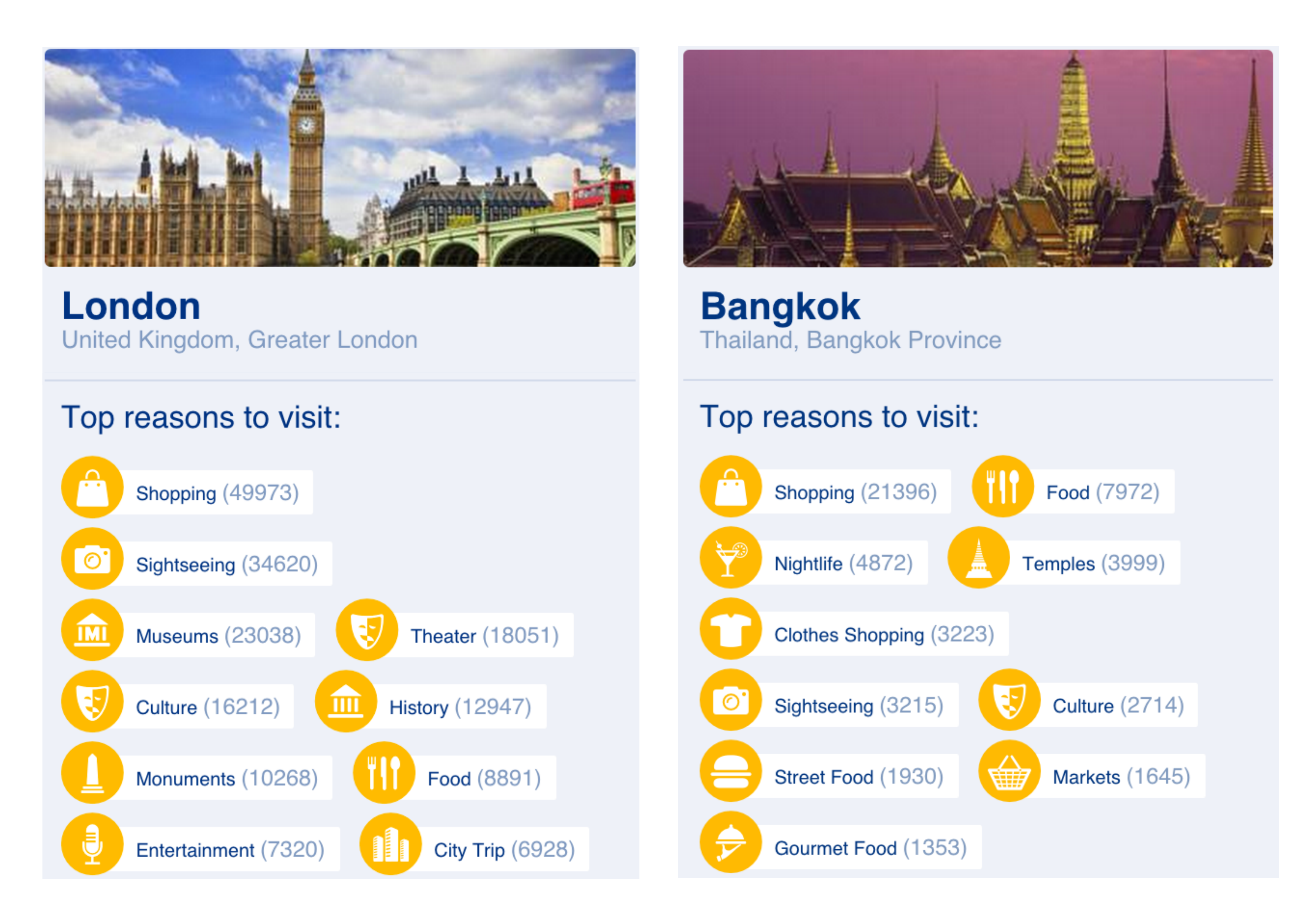}}
\caption{The \df endorsement pages of London and Bangkok.}
\label{fourfig:endors_example}
\end{figure}

The underlying data is based on `endorsements' of users that have booked a hotel at some destination via the online travel agent in the past.  After the users visited the destination, they are asked to endorse the place using a set of endorsements. Initially, the set of endorsements was extracted from users' free-text reviews using a topic-modeling technique such as LDA~\citep{blei_2003, noulas_2014}.  Nowadays, the set of endorsements consists of 256 activities such as `Beach,'  `Nightlife,' `Shopping,' etc.
These endorsements imply that a user liked a destination for particular characteristics.
Two examples of the collected sets of endorsements for two destinations `Bangkok' and `London' are shown in Figure~\ref{fourfig:endors_example}.
As an example of the multi-criteria endorsement data, consider three endorsements: $e_1$ = \emph{`Beach'}, $e_{2}$ = \emph{`Shopping'}, and $e_{3}$ = \emph{`Family Friendly'} and assume that a user $u_j$ after visiting a destination $d_k$ (e.g.`London') provides the review $r_i(u_j,d_k)$ as:
\begin{equation}
\label{eq:review}
r_i(u_j,d_k)=(0,1,0).
\end{equation}
This means our user ranks London for the `Shopping' activity only. However, we cannot conclude that London is not `Family Friendly', i.e.\ negative user opinions are hidden. In contrast to the ratings data of the traditional recommender systems setup, we are dealing with multi-criteria ranking data.
\df is a good example of the service which is working under the \cocos settings from both sides: users and items. 

\mypar{\ucocos at \df}
It is used to plan holidays, so many users visit it infrequently because they have only one or two vacations per year, leading to the sparsity problem. Since users interact with service rarely\:---\:many changes can happen and they might shift their preferences from backpacking activities to family friendly places. Users can use different devices to search over \df without login to the system, so user matching is an actual problem. Users can express different types of preference while planning trips, e.g. they might go to a family friendly resort while traveling with children and look for `Shark Diving' while planning holidays alone, so we need to deal with different user `personas'.

%

\mypar{\icocos at \df}
The list of destinations is growing continuously over time because users share their experience about new places, so we run into the item sparsity problem.
User reviews for destination depends on contextual information. For example, the resort `The Hague at North Sea' is widely endorsed for the activity `Beach' during summer, but not during winter, so we run into the item volatility. Moreover, destination might change over time, e.g. a new aquarium is build and users start to endorse a place for it. Some destinations have different `personas' in which they target several user groups, such as a destination which can be family friendly but at the same time has rich night live. Therefore, we have places that are expressing different `personas'.

These aspects of \cocos at \df can be addressed partially by taking context into account. We propose that the described multi-criteria endorsements can be enhanced by contextual information. We build a contextual ranker for recommending destinations, whereas the current live systems uses an advanced non-contextualized ranker.

\medskip
To summarize, we introduced the continuous cold start problem, and characterized the user and item sides of the \cocos.  We also introduced the \df setup that we used in this paper: \textbf{(1)} we have a set of geographical destinations such as `Paris', `London', `Amsterdam' etc.; \textbf{(2)} each destination is ranked by users who visited the place using a set of endorsements under some situation (which can be described by a set of contexts). In the setting of \cocos, our main goal is to find ways to map any incoming user, without assuming prior history or explicit profiles, to some cluster of like-minded previous users using only contextual data.  In the next section, we will discuss how to discover such contextual user profiles.

%% file: four-sec/3-general_approach.tex
\begin{figure*}[!tb]
\centerline{\includegraphics[width=0.8\textwidth]
{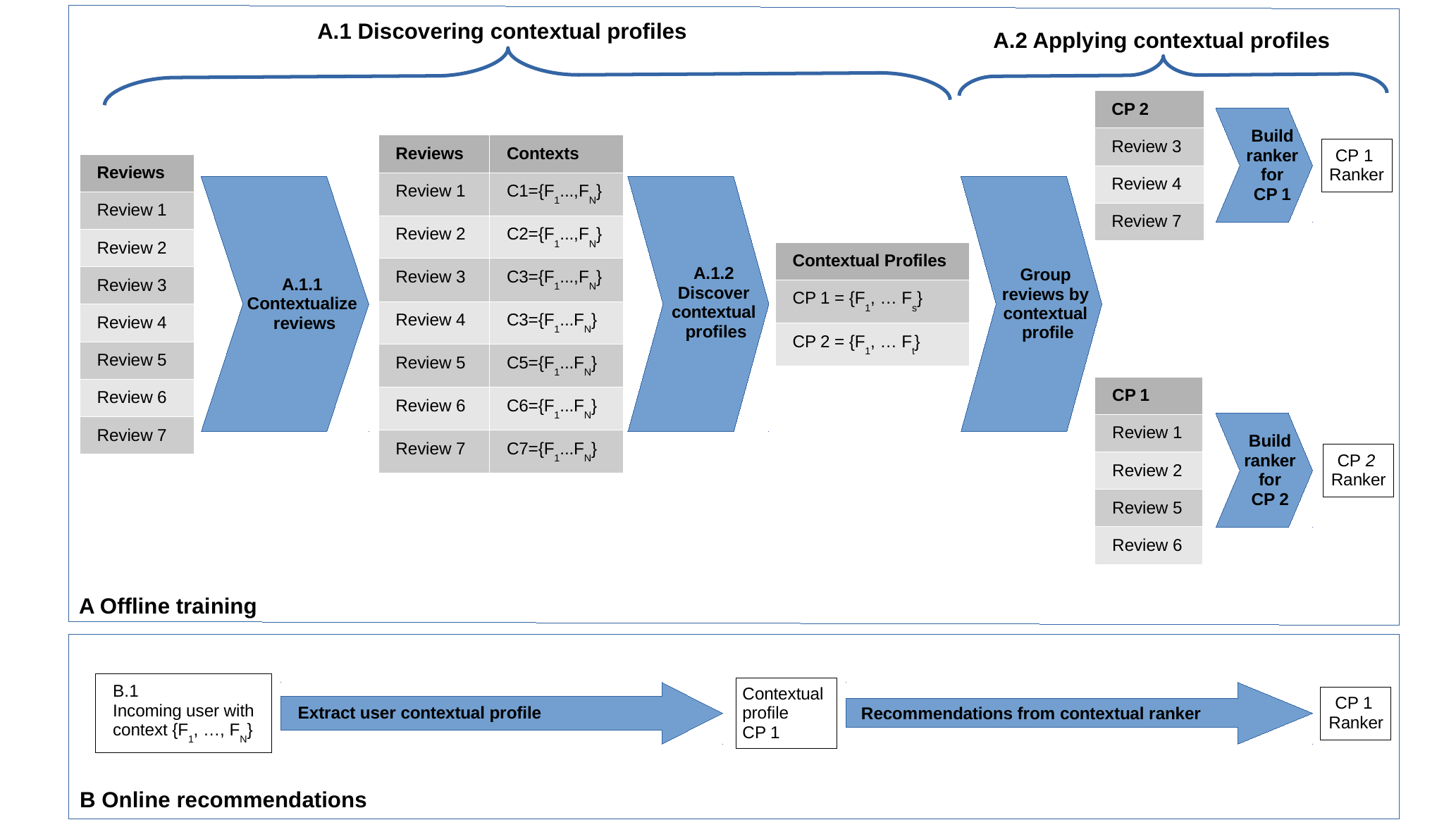}}
\caption{An overall framework for discovering multidimensional contextual user profiles. 
}
\label{fourfig:overall}
\end{figure*}

\section{Multidimensional Contextual \\ User Profiles}
\label{foursec:framework}


In this section we will study our \textsl{\fourrqtwo}
We present an overview of our framework for discovering multidimensional contextual user profiles (CUPs), as outlined in Figure~\ref{fourfig:overall}. It has two main stages: \emph{offline} (\textbf{A}), and \emph{online} (\textbf{B}). 
The discovery of multidimensional CUPs (\textbf{A.1}) happens during the offline stage and is described in Section~\ref{foursec:sit_profiles}. 
The process of using discovered CUPs is as follows:
(\textbf{A.2}) during the \emph{offline stage}, we apply the set of discovered CUPs to learn a customized ranker; and
(\textbf{B}) during the \emph{online stage}, we assign incoming users to one of the CUPs.
The process of using CUPs is presented in Section~\ref{foursec:util_sit_prof}.
Section~\ref{foursec:framework} defines CUPs in a generic way. In Section~\ref{foursec:apply_framework} we show how the framework can be applied to the \df.

\subsection{Defining Contextual User Profiles}
\label{foursec:sit_profiles}

Apart from the reviews, as defined in Equation~\ref{eq:review}, there is additional contextual information about the \emph{situation} in which users made their choice (to consider or not to consider the suggested destination), e.g.\ the geographical location, the time (when a user is using \df), the users' device type, or the referral (where is a user coming from).  We adopt the definition of the context as described in Equation~\ref{eq:context}.

In many real world RS it is not feasible to track user identity information $u_j$ for several reasons: 
\textbf{(1)} privacy issues: only a limited part of the user interaction history can be stored; 
\textbf{(2)} the cold-start problem: when a new user comes without prior history of interaction with the system;
\textbf{(3)} a user does not have to be logged in: so we cannot make use of his interaction history.  
However, we would like to predict user preferences in order to supply him with suitable recommendations.
Therefore, we want to detect a list of typical user situations using contextual information.  This type of situational information we call CUPs.

\begin{figure}[!tb]
\centerline{%
\includegraphics[width=0.45\textwidth]
{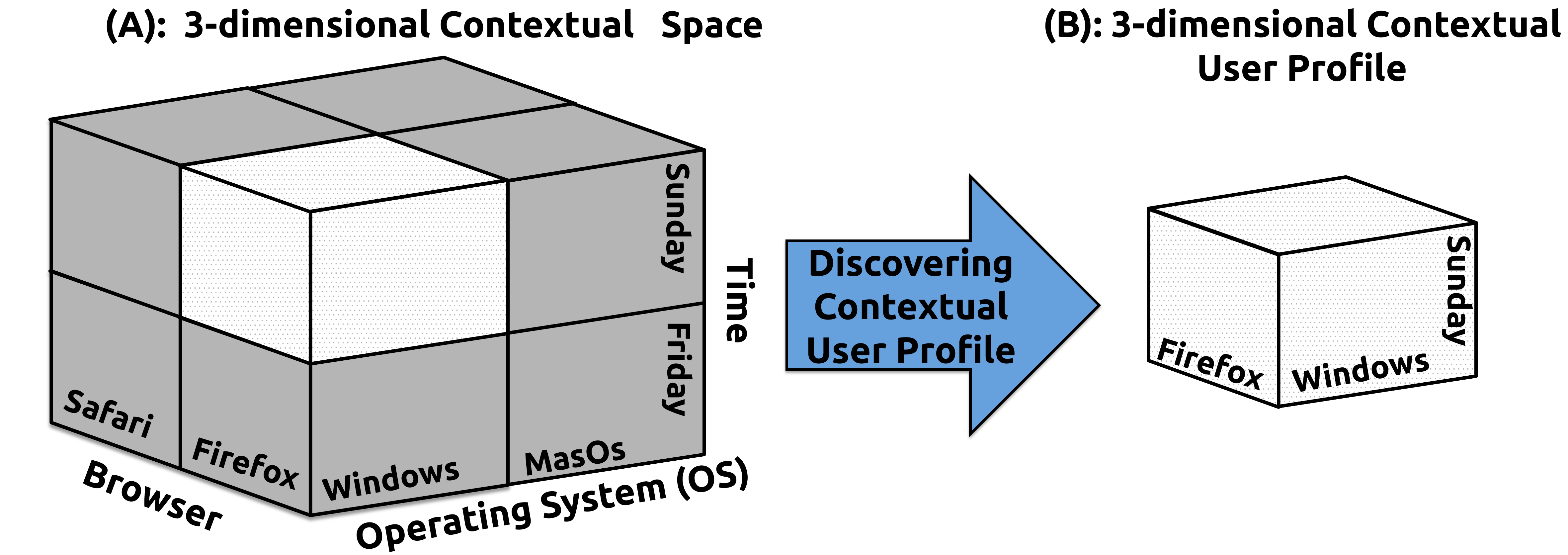}}
\caption{An example for discovering a contextual user profile from $3$-dimensional contextual space. The $3D$ contextual space can be visualized as a cube (A), of which the  contextual user profile is a cube region (B).
}
\label{fourfig:context_space}
\end{figure}

Contextual information can be represented as a $n$-\:dimensional space where the dimensions are the set of contextual features, $\{F_n\}_{n=1}^N$, and the coordinates for each dimension are the contextual categories, $\{v_m\}_{m=1}^M$. 
For example, the contextual feature $F_1$, 'User Device', is represented by the following contextual categories:
\begin{equation}
\label{eq:context_feature}
F_1 = \{v_1 = \text{`Mobile'}, v_2 = \text{`Tablet'}, v_3 = \text{`PC'}\}.
\end{equation}
To simplify the notation we rewrite Equation~\ref{eq:context_feature} as:
\begin{equation}
\label{eq:simp_context_feature}
F_1 = \{F_{11} = \text{`Mobile'}, F_{12} = \text{`Tablet'}, F_{13} = \text{`PC'}\}.
\end{equation}
The $3$-dimensional example (cube) of contextual space is presented in Figure~\ref{fourfig:context_space} (A) where we have three dimensions: $\{F_1=\text{`OS'}, F_2=\text{`Browser'}, F_3=\text{`Time'} \}$.


A \emph{contextual user profile} is a \emph{region} in the $n$-dimensional contextual space that represents `typical' user behavior.
When a user visits our service we can map him to one of the CUPs and use this insight into his preferences to improve the quality of the service, i.e.\ serving better travel recommendations in the \df.

\subsection{Discovering Contextual User Profiles}

We now discuss in more detail the process of discovering CUPs, as outlined in Figure~\ref{fourfig:overall} (A.1). 
The review entities as defined in Equation~\ref{eq:review} can be contextualized, i.e., extended by multidimensional contextual information $C$ as depicted in Figure~\ref{fourfig:overall} (A.1.1). We use the context definition presented in Equation~\ref{eq:simp_context_feature}. The contextual review $r_i$ has the following form: 
\begin{equation}
\label{eq:context_review}
r_i(u_j,d_k)=(e_1,\dots,e_X,F_{11},\dots,F_{NM}),
\end{equation}
where:
\begin{enumerate}
\item $u_j$ is user information that is not stored explicitly, but in our setup we have contextual information regarding how a review is made;
\item $d_k$ is a destination which a user $u_j$ ranks using multi-criteria endorsements;
\item $e_1,\dots,e_X$ are endorsements represented as binary values;
\item $F_{11},\dots,F_{NM}$ are contextual features represented in a binary way.  For example, if a user is using a device with `Windows' as OS and a `Firefox' browser on Sunday, then the context vector is $(1,0,0,1,0,1)$.
\end{enumerate}
In our setup we combine CARS and MCRS presented in Equation~\ref{eq:cars} and~\ref{eq:mcrs} accordingly. A key difference to standard settings is that we are dealing with sparse multi-criteria ranking data, not with ratings. Therefore, negative user opinions are hidden from us. 

Our assumption is that users give similar endorsements in similar situations, and that we can represent it by a subspace of contexts.
In order to enrich the contextual space, we use the review entities with endorsements as an additional dimension to the $n$-dimensional contextual space. 
Some technique can be applied to discover \emph{`closely connected'} regions in the contextual space. After finding the contextual regions in the extended $(n+1)$-dimensional cube we eliminate the endorsement dimension in order to derive CUPs which consist solely of contexts. This allows us to map new incoming users to CUPs.  

The CUP is represented as an agglomeration of a discovered region. For example, if a clustering technique is applied then a cluster center would be an example of CUP, as we will explain in Section~\ref{foursec:clustering}.
In the example in Figure~\ref{fourfig:overall} (A.1.2), we discover two CUPs: $CP_p$ and $CP_q$.
The choice of the clustering method depends on the type of application. We detail the application to the \df in Section~\ref{foursec:clustering}. 
Next, we discuss how the discovered CUPs can be used for ranking suggested destinations.

\subsection{Using Contextual User Profiles}
\label{foursec:util_sit_prof}

The process of using discovered CUPs can be divided into two main parts, see Figure~\ref{fourfig:overall}: 
\textbf{(A.2)} \emph{offline} application of CUPs; and 
\textbf{(B)} \emph{online} mapping of an incoming user to one of the CUPs.

During the offline stage, the set of CUPs can be used for splitting reviews in order to build a set of contextual rankers $\{R_l\}_{l=1}^L$ where $L$ is the number of discovered CUPs.
Our assumption is that a set of contextual rankers serves `better' (more suitable) results than a base ranker $R_b$ which is trained based on all reviews.

During the online stage, an incoming user is mapped to one of the CUPs. A user is represented by a vector of contexts as shown in Figure~\ref{fourfig:overall} (B.1). In order to map a user to one of the CUPs, $CS_1$ or $CS_2$, we can employ any distance metric $D$. The user would be assigned to the \emph{`closest'} CUP, which is $CP_1$ in our example in Figure~\ref{fourfig:overall} (B). Then the user is supplied with a contextual ranker $R_1$ which corresponds to $CS_1$.

\medskip
To summarize, we presented a general framework for discovering and using contextual user profiles.
In principle, any contextual features can be used, including relatively shallow implicit situational context available in any online context.
Also any ratings, reviews or other multi-criteria ranking data can be used, including travel endorsements.  In the next section, we apply the framework to the \df application described in 
Section~\ref{foursec:problem_setup}. 


%% file: four-sec/4-applied_approach.tex
\section{Contextual Travel Recommendations}
\label{foursec:apply_framework}

In this section we will study our \textsl{\fourrqthree}
We present an example how our framework for discovering contextual user profiles (CUPs) from Section~\ref{foursec:framework} can be applied to the \df. 
First, we describe the data used for our experimental pipeline in Section~\ref{foursec:data}.
Second, we use a clustering technique to discover contextual user profiles (CUPs) in Section~\ref{foursec:clustering}. Third, we present in Section~\ref{foursec:util}: 
\textbf{(1)} how these CUPs can be used within a ranking technique based on Naive Bayes;
\textbf{(2)} how the customized rankings are deployed for online user traffic.
We use standard clustering and ranking methods, such as k-means and Naive Bayes, which scale well to the volume of data available. These methods are sufficient to answer our main question about the value of context-aware recommendations. Further optimization is left for future work.

\subsection{Data}
\label{foursec:data}

In the offline training stage, we use reviews collected within the year 2014. The final set contains in total 5,138,494 reviews. We derive two types of data from web logs as contextual information: 
\begin{itemize}
\item \textbf{user agent data} which is presented by \emph{four dimensions} such as \emph{`Device Type'} with $5$ contextual categories (\emph{mobile, tablet} etc.), \emph{`OS'} with $27$ contextual categories (\emph{Windows 8.1, Android, Linux, OS X} etc.), \emph{`Browser'} with $114$ contextual categories (\emph{Internet Explorer 6, Firefox 30, Firefox 34, Safari 7} etc.), and \emph{`Traffic Type'} with 16 contextual categories (\emph{web, mobile browser, application} etc.);
\item \textbf{time data} which is \emph{one dimensional}: the day of the week (\emph{Monday, Tuesday} etc.).
\end{itemize}
This type of contextual information is available in all typical web logs, and can be used to contextualize the reviews as presented in Figure~\ref{fourfig:overall} (A.1.1).
In total, the contextual space has 5 dimensions with 397 coordinates. 
In the online testing stage, we run our experiment on live user traffic for  26,868 users.

\subsection{Clustering Contextualized Reviews}
\label{foursec:clustering}

We use a clustering technique  to discover CUPs as shown in Figure~\ref{fourfig:overall} (A.1.2). We apply k-means clustering~\citep{Jancey_1966} over the set of contextualized reviews as presented in Equation~\ref{eq:context_review}. 
The number of clusters is selected based on Silhouette validation~\citep{Rousseeuw_1987}, which results in 20 clusters as the optimal number.


After obtaining the final set of clusters, we eliminate the endorsement dimension by projecting on the contextual space. We analyze the set of contexts that is associated with the clusters in order to derive the set of CUPs. Because of the projection on the contextual space, clusters may overlap in some contextual categories. 


The cluster centers represent the set of discovered CUPs. We calculate weights for the coordinates of the cluster centres as the ratio of the \emph{(number of times the coordinate $F_{nm}$ appears within cluster $C_i$)} divided by the \emph{(number of times the coordinate $F_{nm}$ appears within all clusters)}. This weight $w_{ij}$ (where $i$ is a cluster identifier and $j$ is an identifier of a coordinate $F_{nm}$) shows how strongly the contextual category $F_{nm}$ is associated with  cluster $i$: The closer $w_{ij}$ to $1$, the stronger the association. 

We employ a pruning technique over the obtained list of CUPs in order to clean up some obvious noise. If $w_{ij}$ is too small for some contextual category $F_{nm}$, then this category is distributed widely over all CUPs and it does not enhance our definition of CUP.
After trails of experiments we empirically determine a threshold:If $F_{nm}$ has $w_{ij}$ $<$ 0.2, we do not include it into the CUPs. 
For example, sometimes contextual categories such as `Monday', `Tuesday' are removed because apparently they do not reflect any `specific' behavior. 
By applying this pruning technique we ended up with 17 clusters. 
%
We present an example of two pruned CUPs in Table~\ref{fourtab:clusters}, which correspond to intuitions about similar users based on context.  It may not be a priori clear why such a cluster provides meaningful context, but the clustering informs us that they have distinct interests and preferences. 

\begin{table}[!tb]
\caption{\strut An example of two obtained cluster centers from real data. Cluster $i$ can be characterized as  `users coming from mobile devices' and Cluster $i+1$  as `users coming from windows-based devices on Fridays and Sundays'.\label{fourtab:clusters}\strut}
\begin{tabularx}{\linewidth}{rXXr}
\toprule
\multicolumn{4}{c}{Cluster}
\\
\multicolumn{1}{c}{i-1}
& \multicolumn{1}{c}{i}
& \multicolumn{1}{c}{i+1}
& \multicolumn{1}{c}{i+2}
\\
\midrule
\ldots
& iPhone.OS.7.Chrome
& Windows.Phone          
& \ldots
\\
& iPhone.OS.5.Chrome
& Windows.Vista
& \\
& iPhone.OS.6.Chrome            
& Friday
& \\
& Android.2.2     
& Sunday
& \\
& Android.2.2.Tablet      
& & \\
& Android.3.1.Tablet      
& & \\
& Android.4.0.Tablet      
& & \\
& Android.4.4.Tablet      
& & \\
& Android.2.1.Tablet      
& & \\
& Android.3.0.Tablet      
& & \\
& Android.4.1    
& & \\
& Android.4.3.Tablet     
& & \\
\bottomrule
\end{tabularx}
\end{table}


Next, we will describe how the discovered CUPs can be applied to destination ranking.

\begin{table*}[!tb]
\caption{Results of the \df A/B testing based on the number of unique users, searches and clicks. The contextual ranker does not significantly change conversion (probability to click at least once), but significantly increases clicks-per-user and click-though-rate (CTR). Significance is assessed as non-overlapping 95\% confidence intervals.\strut}
\vspace{5pt}
\label{fourtab:res}
\centering
\begin{tabular}{lcccccc}
\toprule
Ranker & Users & Searches & Clicks & Conversion  & Clicks/user & CTR\\
\midrule
Baseline 
& 13,306 & 34,463 & 6,373 & 21.7$\pm$0.7\%  & 0.479$\pm$0.012 & 18.5$\pm$0.4\%\\
Contextual
& 13,562 & 35,505 & 7,866 & 21.3$\pm$0.7\% & 0.580$\pm$0.013  & 22.2$\pm$0.4\%\\
\bottomrule
\end{tabular}
\vspace{-10pt}
\end{table*}

\subsection{Using Contextual User Profiles for Destination Ranking}
\label{foursec:util}
As a primary ranking technique we use a Naive Bayes approach. We will describe its application with an example.
Let us consider a user running the searching for `Beach'.  We need to return a ranked list of destinations. For instance, the ranking score for the destination `Miami' is calculated using the following formula:		
\begin{equation}
\label{eq:base_ranker}
P(\text{Miami}, \text{Beach}) =  P (\text{Miami}) \times P(\text{Beach} | \text{Miami});
\end{equation}
where $P(\text{Beach} | \text{Miami})$ is the probability that  the destination Miami gets the endorsement `Beach'. $P (\text{Miami})$ is a prior knowledge about Miami. In the simplistic case the prior would be a ratio of the number of endorsements for Miami to the total number of endorsements in our database. 

If a user uses a second endorsement (e.g.\ + `Food') the ranking score is calculated in the following way:
\begin{equation}
\label{eq:base_ranker_2}
\begin{split}
P(\text{Miami}, \text{Beach}, \text{Food}) = 
P (\text{Miami})  \times P(\text{Beach} | \text{Miami}) \\ 
                  \times  P(\text{Food} | \text{Miami});
\end{split}
\end{equation}
If our user provides $n$ endorsements, Equation~\ref{eq:base_ranker_2} becomes a standard Naive Bayes formula.


We split our set of reviews according to the obtained clusters. Then we train a set of contextual rankers using the same approach as described in Equation~\ref{eq:base_ranker_2} to obtain the customized rankers ${R(C_i)}_{i= 1}^{17}$. This process can be mapped to the general framework presented in Figure~\ref{fourfig:overall} (A.2).

During the online stage, which is shown at general framework work-flow in Figure~\ref{fourfig:overall} (B), an incoming user to the \df is mapped to the closest CUP.  As we use only situational context that does not change per session, we only have to assign our user to the nearest cluster once, and there is no need to update the assignment during the session.  
Then we use a ranker $R(C_i)$ which corresponds to CUP. 

As a distance metric we use Euclidean distance, which deals well with the different nature of some of the clusters (e.g., some clusters capture aspects of the day of the week, and others capture aspects of the used devices).  More advanced mapping of users as mixtures of CUPs is left to future work, as our main goal in this paper is to determine the impact of contextual ranking.



\medskip
To summarize, we described the use of the framework for discovering contextual user profiles for the \df. We contextualized reviews with user agent and time data. Our main goal is to determine the impact of contextual ranking, hence we use standard clustering and ranking methods.  Specifically, we use k-means for clustering and Naive Bayes for ranking and we map incoming users to the nearest cluster based on euclidean distance. 
In the next section, we will present our experimental pipeline which involves online A/B testing at a major travel agent. 

%% file: four-sec/5-experimentation.tex

\section{Experiments and Results}
\label{foursec:experiment}
In this section, we will study our \textsl{{\fourrqfour}}
To test the effectiveness of contextualization, we perform experiments on users of \booking where an instance of the \df is running.  

\subsection{Research Methodology}

We take advantage of a production A/B testing environment at a major online travel agency. A/B testing randomly splits users to see either the baseline or the new variant version of the website, which allows to measure the impact of the new version directly on real users~\citep{tang_kdd_2010,kohavi_kdd_2014}.  As baseline we use a non-contextualized ranker corresponding to the live system. This is an optimized system, trained on a massive volume of traffic, and far superior to standard baselines such as popularity \citep{Kiseleva_sigir_industry_2015xxx}.

As our primary \emph{evaluation metric} in the A/B test, we use clicks-per-user and click-through-rate (CTR)~\citep{lalmas_2014}.
As explained in the motivation, we are dealing with an exploratory task and therefore aim to increase customer engagement.  More clicks-per-user and higher CTR are signals that users click more on the suggested destinations and interact more with the system. 

\subsection{Results}


Table~\ref{fourtab:res} shows the results of our A/B test. We see that the contextual ranker does not significantly change conversion compared to the baseline non-contextual ranker, i.e.\ the probability for a user to click at least once remains the same. Thus, our recommendations do not influence the basic user intent of using the \df. In contrast, the contextual ranker significantly increases further user engagement after the first click: The CTR increases by absolute 3.7\%, and both CTR and clicks-per-user increase dramatically by relative 20\% and 23\%, respectively. Our contextual recommendations invite users to perform more searches and click on more recommendations, both per search and per user. In total, users are significantly more engaged with the \df when presented with contextual recommendations.  

We achieved this substantial increase in clicks with a simple contextualization using straightforward k-means clustering of reviews and a Naive Bayes ranker.  Most computations can be done offline, and only simple calculations have to be performed online. Thus, our model could be trained on large data within reasonable time, and did not negatively impact wallclock and CPU time for the \df web pages in the online A/B test. This is crucial for a webscale production environment \citep{kohavi_kdd_2014}.

\medskip
To summarize, we compared our contextual travel recommendations against the same non-contextualized ranker. This allowed us to compare the effect of contextualization independently of the underlying ranking.
This is a hard baseline corresponding to the current live system applied to the exact same data. 
We observe a dramatic increase in user engagement, with click-through rates and clicks by users increasing by ~20\%.  The simplicity of our contextual models enables us to achieve this engagement without significantly increasing online CPU and memory usage. The experiments clearly demonstrate the value of contextual profiles in a real world application.

%% file: sigir16_cocos.bbl
\begin{thebibliography}{63}
\providecommand{\natexlab}[1]{#1}
\providecommand{\url}[1]{\texttt{#1}}
\expandafter\ifx\csname urlstyle\endcsname\relax
  \providecommand{\doi}[1]{doi: #1}\else
  \providecommand{\doi}{doi: \begingroup \urlstyle{rm}\Url}\fi

\bibitem[Adomavicius and Kwon(2007)]{Adomavicius_2007}
G.~Adomavicius and Y.~Kwon.
\newblock New recommendation techniques for multicriteria rating systems.
\newblock \emph{EXPERT}, 22\penalty0 (3):\penalty0 48--55, 2007.

\bibitem[Adomavicius and Tuzhilin(2005)]{AdomaviciusTuzhilin05}
G.~Adomavicius and A.~Tuzhilin.
\newblock Toward the next generation of recommender systems: a survey of the
  state-of-the-art and possible extensions.
\newblock \emph{TKDE}, 17:\penalty0 734--749, 2005.

\bibitem[Adomavicius and Tuzhilin(2011)]{gediminasadomavicius2010}
G.~Adomavicius and A.~Tuzhilin.
\newblock Context-aware recommender systems.
\newblock In \emph{Recommender Systems Handbook}, pages 217--253, 2011.

\bibitem[Adomavicius et~al.(2005)Adomavicius, Sankaranarayanan, Sen, and
  Tuzhilin]{Adomavicius_2005}
G.~Adomavicius, R.~Sankaranarayanan, S.~Sen, and A.~Tuzhilin.
\newblock Incorporating contextual information in recommender systems using a
  multidimensional approach.
\newblock \emph{TOIS}, 23\penalty0 (1):\penalty0 103--145, 2005.

\bibitem[Adomavicius et~al.(2011)Adomavicius, Manouselis, and
  Kwon]{Adomavicius_2010}
G.~Adomavicius, N.~Manouselis, and Y.~Kwon.
\newblock \emph{Multi-Criteria Recommender Systems}, volume 768-803.
\newblock Recommender Systems Handbook, Springer, 2011.

\bibitem[Agarwal and Chen(2009)]{Agarwal_kdd_2010}
D.~Agarwal and B.-C. Chen.
\newblock Regression-based latent factor models.
\newblock In \emph{KDD}, pages 19--28, 2009.

\bibitem[Agarwal and Chen(2010)]{Agarwal_wsdm_2010}
D.~Agarwal and B.-C. Chen.
\newblock flda: matrix factorization through latent dirichlet allocation.
\newblock In \emph{WSDM}, pages 91--100, 2010.

\bibitem[Anonymous()]{Kiseleva_sigir_industry_2015xxx}
Anonymous.
\newblock Citation withhold to preserve anonymity.
\newblock In \emph{SIGIR}, 2015.

\bibitem[Baltrunas and Ricci(2009{\natexlab{a}})]{Baltrunas_2009}
L.~Baltrunas and F.~Ricci.
\newblock Context-based splitting of item ratings in collaborative filtering.
\newblock In \emph{RecSys}, pages 245--248, 2009{\natexlab{a}}.

\bibitem[Baltrunas and Ricci(2014)]{Baltrunas_2014}
L.~Baltrunas and F.~Ricci.
\newblock Experimental evaluation of context-dependent collaborative filtering
  using item splitting.
\newblock \emph{User Modeling and User-Adapted Interaction}, 24:\penalty0
  7--34, 2014.

\bibitem[Baltrunas and Ricci(2009{\natexlab{b}})]{Baltrunas_cars_2009}
L.~Baltrunas and F.~Ricci.
\newblock Context-dependent items generation in collaborative filtering.
\newblock In \emph{CARS}, 2009{\natexlab{b}}.

\bibitem[Bao et~al.(2012)Bao, Cao, Chen, Tian, and Xiong]{Bao_2012}
T.~Bao, H.~Cao, E.~Chen, J.~Tian, and H.~Xiong.
\newblock An unsupervised approach to modeling personalized contexts of mobile
  users.
\newblock \emph{Knowl. Inf. Syst.}, 31\penalty0 (2):\penalty0 345--370, 2012.

\bibitem[Bel{\'e}m et~al.(2013)Bel{\'e}m, Santos, Almeida, and Gon{\c
  c}alves]{Santos_2013}
F.~Bel{\'e}m, R.~L.~T. Santos, J.~M. Almeida, and M.~A. Gon{\c c}alves.
\newblock Topic diversity in tag recommendation.
\newblock In \emph{RecSys}, pages 141--148, 2013.

\bibitem[Bernardi et~al.(2015)Bernardi, Kamps, Kiseleva, and
  M{\"{u}}ller]{bern_cont15}
L.~Bernardi, J.~Kamps, J.~Kiseleva, and M.~J.~I. M{\"{u}}ller.
\newblock The continuous cold-start problem in e-commerce recommender systems.
\newblock In \emph{CBRecSys}, pages 30--33, 2015.

\bibitem[Blei et~al.(2003)Blei, Ng, and Jordan]{blei_2003}
D.~M. Blei, A.~Y. Ng, and M.~I. Jordan.
\newblock Latent dirichlet allocation.
\newblock \emph{JMLR}, 3:\penalty0 993--1022, 2003.

\bibitem[Cao et~al.(2010)Cao, Bao, Yang, Chen, and Tian]{Cao_cikm_2010}
H.~Cao, T.~Bao, Q.~Yang, E.~Chen, and J.~Tian.
\newblock An effective approach for mining mobile user habits.
\newblock In \emph{CIKM}, pages 1677--1680, 2010.

\bibitem[Census Bureau()]{estat:url}
Census Bureau.
\newblock E-stats: Measuring the electronic economy.
\newblock https://www.census.gov/econ/estats/, 2015.

\bibitem[Chua et~al.(2013)Chua, Oentaryo, and Lim]{chua2013modeling}
F.~C.~T. Chua, R.~J. Oentaryo, and E.-P. Lim.
\newblock Modeling temporal adoptions using dynamic matrix factorization.
\newblock In \emph{ICDM}, pages 91--100, 2013.

\bibitem[Dong et~al.(2010)Dong, Chang, Zheng, Mishne, Bai, Zhang, Buchner,
  Liao, and Diaz]{dong:towa10}
A.~Dong, Y.~Chang, Z.~Zheng, G.~Mishne, J.~Bai, R.~Zhang, K.~Buchner, C.~Liao,
  and F.~Diaz.
\newblock Towards recency ranking in web search.
\newblock In \emph{WSDM}, pages 11--20, 2010.

\bibitem[Hariri et~al.(2012)Hariri, Mobasher, and Burke]{Hariri_2012}
N.~Hariri, B.~Mobasher, and R.~D. Burke.
\newblock Context-aware music recommendation based on latent topic sequential
  patterns.
\newblock In \emph{RecSys}, pages 131--138, 2012.

\bibitem[Hariri et~al.(2013)Hariri, Mobasher, and Burke]{Hariri_2013}
N.~Hariri, B.~Mobasher, and R.~D. Burke.
\newblock Query-driven context aware recommendation.
\newblock In \emph{RecSys}, pages 9--16, 2013.

\bibitem[Hawalah and Fasli(2014)]{Hawalah_2014}
A.~Hawalah and M.~Fasli.
\newblock Utilizing contextual ontological user profiles for personalized
  recommendations.
\newblock \emph{Expert Syst. Appl. (ESWA)}, 41\penalty0 (10):\penalty0
  4777--4797, 2014.

\bibitem[Ho et~al.(2014)Ho, Chiang, and Hsu Yung-Jen]{Ho_wsdm_2014}
Y.-C. Ho, Y.-T. Chiang, and J.~Hsu Yung-Jen.
\newblock Who likes it more?: mining worth-recommending items from long tails
  by modeling relative preference.
\newblock In \emph{WSDM}, pages 253--262, 2014.

\bibitem[Hu et~al.(2014)Hu, Hall, and Attenberg]{Hu:2014:SLT:2623330.2623338}
D.~J. Hu, R.~Hall, and J.~Attenberg.
\newblock Style in the long tail: Discovering unique interests with latent
  variable models in large scale social e-commerce.
\newblock In \emph{KDD}, pages 1640--1649, 2014.

\bibitem[Jancey(1966)]{Jancey_1966}
R.~Jancey.
\newblock Multidimensional group analysis.
\newblock \emph{Australian Journal of Botany}, 14\penalty0 (1):\penalty0
  127--130, 1966.

\bibitem[Jannach et~al.(2012)Jannach, Karakaya, and Gedikli]{Jannach_2012}
D.~Jannach, Z.~Karakaya, and F.~Gedikli.
\newblock Accuracy improvements for multi-criteria recommender system.
\newblock In \emph{EC}, pages 674--689, 2012.

\bibitem[Kapoor et~al.(2015)Kapoor, Subbian, Srivastava, and
  Schrater]{Kapoor_wsdm_2015}
K.~Kapoor, K.~Subbian, J.~Srivastava, and P.~Schrater.
\newblock Just in time recommendations: Modeling the dynamics of boredom in
  activity streams.
\newblock In \emph{WSDM}, pages 233--242, 2015.

\bibitem[Karatzoglou et~al.(2010)Karatzoglou, Amatriain, Baltrunas, and
  Oliver]{Karatzoglou_2010}
A.~Karatzoglou, X.~Amatriain, L.~Baltrunas, and N.~Oliver.
\newblock Multiverse recommendation: n-dimensional tensor factorization for
  context-aware collaborative filtering.
\newblock In \emph{RecSys}, pages 79--86, 2010.

\bibitem[Kiseleva et~al.(2013{\natexlab{a}})Kiseleva, Lam, Pechenizkiy, and
  Calders]{kiseleva_dddm_2013}
J.~Kiseleva, H.~T. Lam, M.~Pechenizkiy, and T.~Calders.
\newblock Predicting current user intent with contextual markov models.
\newblock In \emph{ICDM Workshops}, 2013{\natexlab{a}}.

\bibitem[Kiseleva et~al.(2013{\natexlab{b}})Kiseleva, Lam, Pechenizkiy, and
  Calders]{kiselevalpc_www13}
J.~Kiseleva, H.~T. Lam, M.~Pechenizkiy, and T.~Calders.
\newblock Discovering temporal hidden contexts in web sessions for user trail
  prediction.
\newblock In \emph{TempWeb@WWW'2013}, pages 1067--1074, 2013{\natexlab{b}}.

\bibitem[Kiseleva et~al.(2015)Kiseleva, M{\"{u}}ller, Bernardi, Davis, Kovacek,
  Einarsen, Kamps, Tuzhilin, and Hiemstra]{Kiseleva_sigir_industry_2015}
J.~Kiseleva, M.~J.~I. M{\"{u}}ller, L.~Bernardi, C.~Davis, I.~Kovacek, M.~S.
  Einarsen, J.~Kamps, A.~Tuzhilin, and D.~Hiemstra.
\newblock Where to go on your next trip?: Optimizing travel destinations based
  on user preferences.
\newblock In \emph{SIGIR (Industry Track)}, pages 1097--1100, 2015.

\bibitem[Kluver and Konstan(2014)]{Kluver:2014:ERB:2645710.2645742}
D.~Kluver and J.~A. Konstan.
\newblock Evaluating recommender behavior for new users.
\newblock In \emph{RecSys}, pages 121--128, 2014.

\bibitem[Kohavi et~al.(2014)Kohavi, Deng, Longbotham, and Xu]{kohavi_kdd_2014}
R.~Kohavi, A.~Deng, R.~Longbotham, and Y.~Xu.
\newblock Seven rules of thumb for web site experimenters.
\newblock In \emph{KDD}, pages 1857--1866, 2014.

\bibitem[Koren(2010)]{koren2010collaborative}
Y.~Koren.
\newblock Collaborative filtering with temporal dynamics.
\newblock \emph{Communications of the ACM}, 53\penalty0 (4):\penalty0 89--97,
  2010.

\bibitem[Lakiotaki et~al.(2011)Lakiotaki, Matsatsinis, and
  Tsoukias]{Lakiotaki_2011}
K.~Lakiotaki, N.~F. Matsatsinis, and A.~Tsoukias.
\newblock Multicriteria user modeling in recommender systems.
\newblock \emph{IEEE Intelligent System}, 26\penalty0 (2):\penalty0 64--76,
  2011.

\bibitem[Lalmas et~al.(2014)Lalmas, O'Brien, and Yom-Tov]{lalmas_2014}
M.~Lalmas, H.~O'Brien, and E.~Yom-Tov.
\newblock Measuring user engagement.
\newblock \emph{Synthesis Lectures on Information Concepts, Retrieval, and
  Services}, 6\penalty0 (4):\penalty0 1--132, 2014.

\bibitem[Liu et~al.(2011)Liu, Mehandjiev, and Xu]{Liu_2011}
L.~Liu, N.~Mehandjiev, and D.-L. Xu.
\newblock Multi-criteria service recommendation based on user criteria
  preferences.
\newblock In \emph{RecSys}, pages 77--84, 2011.

\bibitem[Ma et~al.(2012)Ma, Cao, Yang, Chen, and Tian]{Ma_www_2012}
H.~Ma, H.~Cao, Q.~Yang, E.~Chen, and J.~Tian.
\newblock A habit mining approach for discovering similar mobile users.
\newblock In \emph{WWW}, pages 231--240, 2012.

\bibitem[Montanez et~al.(2014)Montanez, White, and Huang]{Montanez_cikm_2015}
G.~D. Montanez, R.~W. White, and X.~Huang.
\newblock Cross-device search.
\newblock In \emph{CIKM}, pages 1669--1678, 2014.

\bibitem[Noulas and Einarsen(2014)]{noulas_2014}
A.~Noulas and M.~S. Einarsen.
\newblock User engagement through topic modelling in travel.
\newblock In \emph{Second Workshop on User Engagement Optimization}, 2014.

\bibitem[Palmisano et~al.(2008)Palmisano, Tuzhilin, and
  Gorgoglione]{palmisano2008}
C.~Palmisano, A.~Tuzhilin, and M.~Gorgoglione.
\newblock Using context to improve predictive modeling of customers in
  personalization applications.
\newblock \emph{TKDE}, 2008, November.

\bibitem[Panniello et~al.(2009)Panniello, Tuzhilin, Gorgoglione, Palmisano, and
  Pedone]{pann:expe09}
U.~Panniello, A.~Tuzhilin, M.~Gorgoglione, C.~Palmisano, and A.~Pedone.
\newblock Experimental comparison of pre- vs. post-filtering approaches in
  context-aware recommender systems.
\newblock In \emph{RecSys}, pages 265--268, 2009.

\bibitem[Rashid et~al.(2002)Rashid, Albert, Cosley, Lam, McNee, Konstan, and
  Riedl]{RashidRiedls02}
A.~M. Rashid, I.~Albert, D.~Cosley, S.~K. Lam, S.~M. McNee, J.~A. Konstan, and
  J.~Riedl.
\newblock Getting to know you: Learning new user preferences in recommender
  systems.
\newblock In \emph{IUI}, pages 127--134, 2002.

\bibitem[Rendle(2009)]{Rendle_2009}
S.~Rendle.
\newblock Factorization machines.
\newblock In \emph{ICDM}, pages 995--1000, 2009.

\bibitem[Rendle et~al.(2011)Rendle, Gantner, Freudenthaler, and
  Schmidt-Thieme]{steffenrendle}
S.~Rendle, Z.~Gantner, C.~Freudenthaler, and L.~Schmidt-Thieme.
\newblock Fast context-aware recommendations with factorization machines.
\newblock In \emph{SIGIR}, pages 635--644, 2011.

\bibitem[Resnick et~al.(1994)Resnick, Iacovou, Suchak, Bergstrom, and
  Riedl]{ResnickRiedl94}
P.~Resnick, N.~Iacovou, M.~Suchak, P.~Bergstrom, and J.~Riedl.
\newblock Grouplens: An open architecture for collaborative filtering of
  netnews.
\newblock In \emph{CSCW}, pages 175--186, 1994.

\bibitem[Rousseeuw(1987)]{Rousseeuw_1987}
P.~J. Rousseeuw.
\newblock Silhouettes: A graphical aid to the interpretation and validation of
  cluster analysis.
\newblock \emph{JCAM}, 20:\penalty0 53--65, 1987.

\bibitem[Saveski and Mantrach(2014)]{SaveskiMantrach14}
M.~Saveski and A.~Mantrach.
\newblock Item cold-start recommendations: Learning local collective
  embeddings.
\newblock In \emph{RecSys}, pages 89--96, 2014.

\bibitem[Schein et~al.(2002)Schein, Popescul, Ungar, and
  Pennock]{schein2002methods}
A.~I. Schein, A.~Popescul, L.~H. Ungar, and D.~M. Pennock.
\newblock Methods and metrics for cold-start recommendations.
\newblock In \emph{SIGIR}, pages 253--260, 2002.

\bibitem[Sedhain et~al.(2014)Sedhain, Sanner, Braziunas, Xie, and
  Christensen]{sedhain2014social}
S.~Sedhain, S.~Sanner, D.~Braziunas, L.~Xie, and J.~Christensen.
\newblock Social collaborative filtering for cold-start recommendations.
\newblock In \emph{RecSys}, pages 345--348, 2014.

\bibitem[Shardanand and Maes(1995)]{ShardanandMaes95}
U.~Shardanand and P.~Maes.
\newblock Social information filtering: Algorithms for automating 'word of
  mouth'.
\newblock In \emph{CHI}, pages 210--217, 1995.

\bibitem[Shi et~al.(2012)Shi, Karatzoglou, Baltrunas, Larson, Hanjalic, and
  Oliver]{Shi_2012}
Y.~Shi, A.~Karatzoglou, L.~Baltrunas, M.~Larson, A.~Hanjalic, and N.~Oliver.
\newblock Tfmap: optimizing map for top-n context-aware recommendation.
\newblock In \emph{SIGIR}, pages 155--164, 2012.

\bibitem[Shi et~al.(2014)Shi, Karatzoglou, Baltrunas, Larson, and
  Hanjalic]{Shi_cikm_2014}
Y.~Shi, A.~Karatzoglou, L.~Baltrunas, M.~Larson, and A.~Hanjalic.
\newblock Cars2: Learning context-aware representations for context-aware
  recommendations.
\newblock In \emph{CIKM}, pages 291--300, 2014.

\bibitem[Sun et~al.(2012)Sun, Varshney, and Subbian]{sun2012dynamic}
J.~Z. Sun, K.~R. Varshney, and K.~Subbian.
\newblock Dynamic matrix factorization: A state space approach.
\newblock In \emph{ICASSP}, pages 1897--1900, 2012.

\bibitem[Tang et~al.(2010)Tang, Agarwal, O'Brien, and Meyer]{tang_kdd_2010}
D.~Tang, A.~Agarwal, D.~O'Brien, and M.~Meyer.
\newblock Overlapping experiment infrastructure: More, better, faster
  experimentation.
\newblock In \emph{KDD}, pages 17--26, 2010.

\bibitem[Tang et~al.(2013)Tang, Gao, Hu, and Liu]{Tang_2013}
J.~Tang, H.~Gao, X.~Hu, and H.~Liu.
\newblock Context-aware review helpfulness rating prediction.
\newblock In \emph{RecSys}, pages 1--8, 2013.

\bibitem[Tang et~al.(2014)Tang, Jiang, Li, and
  Li]{Tang:2014:ECB:2645710.2645732}
L.~Tang, Y.~Jiang, L.~Li, and T.~Li.
\newblock Ensemble contextual bandits for personalized recommendation.
\newblock In \emph{RecSys}, pages 73--80, 2014.

\bibitem[Tavakol and Brefeld(2014)]{Tavakol:2014:FMD:2645710.2645739}
M.~Tavakol and U.~Brefeld.
\newblock Factored mdps for detecting topics of user sessions.
\newblock In \emph{RecSys}, pages 33--40, 2014.

\bibitem[TREC()]{trec:url}
TREC.
\newblock Text retrieval conference: Contextual suggestion track.
\newblock {https://sites.google.com/site/treccontext/}, 2015.

\bibitem[Turney(2002)]{turny2002_discovery}
P.~Turney.
\newblock The identification of context-sensitive features: A formal definition
  of context for concept learning.
\newblock \emph{CoRR}, 2002.

\bibitem[Vargas-Govea et~al.(2011)Vargas-Govea, Gonz{\'a}lez-Serna, and
  Ponce-Medell{\'\i}n]{Blanca_2011}
B.~Vargas-Govea, J.~G. Gonz{\'a}lez-Serna, and R.~Ponce-Medell{\'\i}n.
\newblock Effects of relevant contextual features in the performance of a
  restaurant recommender system.
\newblock In \emph{CARS}, 2011.

\bibitem[White and Dumais(2009)]{whit:char09}
R.~W. White and S.~T. Dumais.
\newblock Characterizing and predicting search engine switching behavior.
\newblock In \emph{CIKM}, pages 87--96, 2009.

\bibitem[Zhu et~al.(2012)Zhu, Cao, Chen, Xiong, and Tian]{Zhu_cikm_2012}
H.~Zhu, H.~Cao, E.~Chen, H.~Xiong, and J.~Tian.
\newblock Exploiting enriched contextual information for mobile app
  classification.
\newblock In \emph{CIKM}, pages 1617--1621, 2012.

\end{thebibliography}
